\documentclass{mn2e}
\usepackage{psfig}

\title[2dFGRS: luminosity functions by density environment and galaxy type]
{The 2dF Galaxy Redshift Survey: luminosity functions by density
  environment and galaxy type}

\author[Croton et~al.]{ \parbox[t]{\textwidth}{ 
Darren J.\ Croton$^1$,
Glennys R.\ Farrar$^2$, 
Peder Norberg$^3$, 
Matthew Colless$^4$, 
John A.\ Peacock$^5$, 
I. K.\ Baldry$^6$,
C. M.\ Baugh$^7$, 
J. Bland-Hawthorn$^4$, 
T. Bridges$^8$,
R. Cannon$^4$, 
S. Cole$^7$, 
C. Collins$^9$, 
W. Couch$^{10}$, 
G. Dalton$^{11,12}$, 
R. De Propris$^{13,14}$, 
S. P.\ Driver$^{13}$, 
G. Efstathiou$^{15}$, 
R. S.\ Ellis$^{16}$,
C. S.\ Frenk$^7$, 
K. Glazebrook$^6$, 
C. Jackson$^{17}$, 
O. Lahav$^{18}$, 
I. Lewis$^{11}$, 
S. Lumsden$^{19}$, 
S. Maddox$^{20}$, 
D. Madgwick$^{21}$, 
B. A.\ Peterson$^{13}$, 
W. Sutherland$^5$,
K. Taylor$^{16}$
(The 2dFGRS Team)
}
  \vspace*{6pt} \\
$^1$Max-Planck-Institut f\"ur Astrophysik, D-85740 Garching, Germany \\
$^2$Center for Cosmology and Particle Physics, Department of Physics,
  New York University, New York NY 10003 \\ 
$^3$ETHZ Institut f\"ur Astronomie, HPF G3.1, ETH H\"onggerberg, CH-8093
       Z\"urich, Switzerland \\
$^4$Anglo-Australian Observatory, P.O.\ Box 296, Epping, NSW 2111,
    Australia\\  
$^5$Institute for Astronomy, University of Edinburgh, Royal Observatory, 
       Blackford Hill, Edinburgh EH9 3HJ, UK.\\ 
$^6$Department of Physics \& Astronomy, Johns Hopkins University,
       Baltimore, MD 21118-2686, USA \\
$^7$Department of Physics, University of Durham, South Road, 
    Durham DH1 3LE, UK \\ 
$^8$Department of Physics, Queen's University, Kingston, 
    Ontario K7L 3N6, Canada \\
$^9$Astrophysics Research Institute, Liverpool John Moores University,  
    Twelve Quays House, Birkenhead, L14 1LD, UK \\
$^{10}$Department of Astrophysics, University of New South Wales, Sydney, 
    NSW 2052, Australia \\
$^{11}$Department of Physics, University of Oxford, Keble Road, 
    Oxford OX1 3RH, UK \\
$^{12}$Space Science \& Technology Division, Rutherford Appleton Laboratory, 
    Chilton OX11 0QX, UK \\
$^{13}$Research School of Astronomy \& Astrophysics, The Australian 
    National University, Weston Creek, ACT 2611, Australia \\
$^{14}$Astrophysics Group, Department of Physics, Bristol University,
    Tyndall Avenue, Bristol, BS8 1TL, UK \\
$^{15}$Institute of Astronomy, University of Cambridge, Madingley Road,
    Cambridge CB3 0HA, UK \\
$^{16}$Department of Astronomy, California Institute of Technology, 
    Pasadena, CA 91025, USA \\
$^{17}$CSIRO Australia Telescope National Facility, PO
    Box 76, Epping, NSW 1710, Australia \\
$^{18}$Department of Physics and Astronomy, University College London, 
    Gower Street, London WC1E 6BT, UK \\
$^{19}$Department of Physics, University of Leeds, Woodhouse Lane,
       Leeds, LS2 9JT, UK \\
$^{20}$School of Physics \& Astronomy, University of Nottingham,
       Nottingham NG7 2RD, UK \\
$^{21}$Department of Astronomy, University of California, Berkeley, 
       CA 94720, USA \\
\vspace{-0.5cm}
}

\date{Accepted ---. Received ---;in original form ---}

\pubyear{2003}

\newcommand{\plotone}[1]
           {\centering \leavevmode \psfig{file=#1,width=\columnwidth,clip=}}

\newcommand{\plotfull}[1]
           {\centering \leavevmode \psfig{file=#1,width=\textwidth,clip=}}
\newcommand{\bJ}{\mbox{{\rm b}$_{\rm J}$}}
\newcommand{\rF}{\mbox{{\rm r}$_{\rm F}$}}

\begin{document}

\maketitle

\begin{abstract}
  We use the 2dF Galaxy Redshift Survey to measure the dependence of
  the \bJ-band galaxy luminosity function on large-scale
  environment, defined by density contrast in spheres of radius
  $8h^{-1}$Mpc, and on spectral type, determined from principal
  component analysis. We find that the galaxy populations at both
  extremes of density differ significantly from that at the mean
  density. 
  The population in voids is dominated by late types and shows, relative
  to the mean, a deficit of galaxies that becomes increasingly
  pronounced at magnitudes brighter than \smash{$M_{\rm
  b_J}-5\log_{10}h \la -18.5$}. In contrast, cluster regions have a
  relative excess of very bright early-type galaxies with
  \smash{$M_{\rm b_J}-5\log_{10}h \la -21$}.  Differences in the mid
  to faint-end population between environments are significant: at
  \smash{$M_{\rm b_J}-5\log_{10}h=-18$} early and late-type cluster
  galaxies show comparable abundances, whereas in voids the late types
  dominate by almost an order of magnitude.
  We find that the luminosity functions measured in all density
  environments, from voids to clusters, can be approximated by
  Schechter functions with parameters that vary smoothly with local
  density, but in a fashion which differs strikingly for early and
  late-type galaxies.  These observed variations, combined with our
  finding that the faint-end slope of the overall luminosity function
  depends at most weakly on density environment, may prove to be a
  significant challenge for models of galaxy formation.
\end{abstract}


\section{Introduction}

The galaxy luminosity function has played a central role in the
development of modern observational and theoretical astrophysics, and is
a well established and fundamental tool for measuring the large-scale
distribution of galaxies in the universe (Efstathiou, Ellis \& Peterson
1988; Loveday et~al.\ 1992; Marzke, Huchra \& Geller 1994; Lin et~al.
1996; Zucca et~al.\ 1997; Ratcliffe et~al.\ 1998;\ Norberg et~al.\
2002a; Blanton et~al.\ 2003a;). 
The galaxy luminosity function of the 2dF Galaxy Redshift Survey
(2dFGRS) has been characterised in several papers: Norberg et al.
(2002a) consider the survey as a whole; Folkes et~al. (1999) and
Madgwick et al. (2002) split the galaxy population by spectral type;
De Propris et al. (2003) measure the galaxy luminosity function of
clusters in the 2dFGRS; Eke et al. (2004) estimate the galaxy
luminosity function for groups of different mass.
Such targeted studies are invaluable if one wishes to understand how galaxy
properties are influenced by external factors such as local density
environment (e.g. the differences between cluster and field galaxies).

A natural extension of such work is to examine a wider range of galaxy
environments and how specific galaxy properties transform as one moves
between them, from the very under-dense `void' regions, to mean
density regions, to the most over-dense `cluster' regions.  In order
to `connect the dots' between galaxy populations of different type and
with different local density a more comprehensive analysis needs to be
undertaken. Although progress has been made in this regard on both the
observational front (Bromley et~al.\ 1998; Christlein 2000; H\"{u}tsi
et~al.\ 2002) and theoretical front (Peebles 2001; Mathis \& White
2002; Benson et~al. 2003; Mo et~al. 2004), past galaxy redshift
surveys have been severely limited in both their small galaxy numbers
and small survey volumes. Only with the recent emergence of large
galaxy redshift surveys such as the 2dFGRS and also the Sloan Digital
Sky Survey (SDSS) can such a study be undertaken with any reasonable
kind of precision (for the SDSS, see recent work by Hogg et al. 2003,
Hoyle et al. 2003, and Kauffmann et al. 2004).

In this paper we use the 2dFGRS galaxy catalogue to provide an
extensive description of the luminosity distribution of galaxies in
the local universe for all density environments within the 2dFGRS
survey volume.  In addition, the extreme under-dense and over-dense
regions of the survey are further dissected as a function of 2dFGRS
galaxy spectral type, $\eta$, which can approximately be cast as early
and late-type galaxy populations (Madgwick et~al.\ 2002, see Section
2).
The void galaxy population is especially interesting as it is only
with these very large survey samples and volumes possible to measure
it with any degree of accuracy.  Questions have been raised (e.g.,
Peebles 2001) as to whether the standard $\Lambda$CDM cosmology
correctly describes voids, most notably in relation to reionisation
and the significance of the dwarf galaxy population in such under-dense
regions.

This paper is organised as follows. In Section~2 we provide a brief
description of the 2dFGRS and the way in which we measure the galaxy
luminosity function from it. The luminosity function results are
presented in Section~3, and then compared with past results in
Section~4.  We discuss the implications for models of galaxy formation in
Section~5.  Throughout we assume a $\Lambda$CDM cosmology with
parameters $\Omega_m=0.3$, $\Omega_{\Lambda}=0.7$, and
$H_0=100h^{-1}$kms$^{-1}$Mpc$^{-1}$. 

\section{METHOD}

\subsection{The 2dFGRS survey}

We use the completed 2dFGRS as our starting point (Colless et~al.\
2003), giving a total of 221,414 high quality redshifts. The median
depth of the full survey, to a nominal magnitude limit of $b_{\rm J}
\approx 19.45$, is $z \approx 0.11$. We consider the two large
contiguous survey regions, one in the south Galactic pole and one
towards the north Galactic pole.  To improve the accuracy of our
measurement our attention is restricted to the parts of the survey
with high redshift completeness ($>70\%$) and galaxies with apparent
magnitude $b_{\rm J} < 19.0$, well within the above survey limit (see
also Appendix~C).  Our conclusions remain unchanged for reasonable
choices of both these restrictions. Full details of the 2dFGRS and the
construction and use of the mask quantifying the completeness of the
survey can be found in Colless et~al.\ (2001, 2003) and Norberg et
al. (2002a).

Where possible, galaxy spectral types are determined using the
principal component analysis (PCA) of Madgwick et~al.\ (2002) and the
classification quantified by a spectral parameter, $\eta$. This allows
us to divide the galaxy sample into two broad classes, conventionally
called late and early types for brevity. The late types are those with
$\eta \ge -1.4$ that have active star formation and the early types
are the more quiescent galaxy population with $\eta <
-1.4$. Approximately $90\%$ of the galaxy catalogue can be classified
in this way. This division at $\eta=-1.4$ corresponds to an obvious
dip in the $\eta$ distribution (Section~2.4; see also Madgwick et~al.\
2002) and a similar feature in the $\bJ-\rF$ colour distribution, and
therefore provides a fairly natural partition between early and late
types. 
When calculating each galaxy's absolute magnitude we apply the
spectral type dependent $k+e$ corrections of Norberg et~al.\ (2002a);
when no type can be measured we use their mean $k+e$ correction. In
this way all galaxy magnitudes have been corrected to zero redshift.

\subsection{Local density measurement}

The 2dFGRS galaxy catalogue is {\em magnitude limited}; it has a fixed
apparent magnitude limit which corresponds to a faint absolute
magnitude limit that becomes brighter at higher redshifts. Over any
given range of redshift there is a certain range of absolute
magnitudes within which all galaxies can be seen by the survey and are
thus included in the catalogue (apart from a modest incompleteness in
obtaining the galaxies' redshifts). Selecting galaxies within these
redshift and absolute magnitude limits defines a {\em volume-limited}
sub-sample of galaxies from the magnitude-limited catalogue (see
e.g. Norberg et~al.\ 2001, 2002b; Croton et~al.\ 2004); this sub-sample is
complete over the specified redshift and absolute magnitude ranges.

To estimate the local density for each galaxy we first need
to establish a volume-limited {\em density defining population} (DDP)
of galaxies.  This population is used to fix the density contours
in the redshift space volume containing the
magnitude-limited galaxy catalogue.  We restrict the magnitude-limited
survey to the redshift range $0.05<z<0.13$, giving an effective
sampling volume of approximately $7 \times 10^6 h^{-3}$Mpc$^3$.  Such
a restriction guarantees that all galaxies in the magnitude range
$-19>M_{\rm b_J}-5\log_{10} h>-22$ (i.e. effectively brighter than
$M^*+0.7$) are volume limited, and allows us to use this
sub-population as the DDP.  The mean number density of DDP galaxies is
$8.6 \times 10^{-3} h^3$Mpc$^{-3}$.  In Appendix~B we consider the
effect of changing the magnitude range of the DDP and find only a very
small difference in our final results.

The local density contrast for each magnitude-limited galaxy is
determined by counting the number of DDP neighbours within an
$8h^{-1}$Mpc radius, $N_g$, and comparing this with the expected
number, $\bar{N_g}$, obtained by integrating under the published
luminosity 2dFGRS function of Norberg et~al. (2002a) over the same
magnitude range that defines the DDP:
\begin{equation}\label{delta8}
\delta_8 \equiv \frac{\delta \rho_g}{\rho_g} =
\frac{N_g-\bar{N_g}}{\bar{N_g}}\ \Bigg{|}_{R=8h^{-1}\rm{Mpc}} ~.
\end{equation}
In Appendix~B we explore the effect of changing this smoothing scale
from between $4h^{-1}$Mpc to $12h^{-1}$Mpc.  We find that our
conclusions remain unchanged, although, not surprisingly,
smaller scale spheres tend to sample the under dense regions
differently.  Spheres of $8h^{-1}$Mpc are found to be the best probe
of both the under and over-dense regions of the survey.

With the above restrictions, the magnitude-limited galaxy sample
considered in our analysis contains a total of $81,387$ ($51,596$)
galaxies brighter than $M_{\rm b_J} -5\log_{10}h = -17~ (-19)$, with
$30,354$ ($23,043$) classified as early types and $42,772$ ($23,815$)
classified as late types.  Approximately $70\%$ of all galaxies in
this sample are sufficiently within the survey boundaries to be given
a local density.  Details of the different sub-samples binned by local
density and type are given in Table~1.

\begin{figure}
\plotone{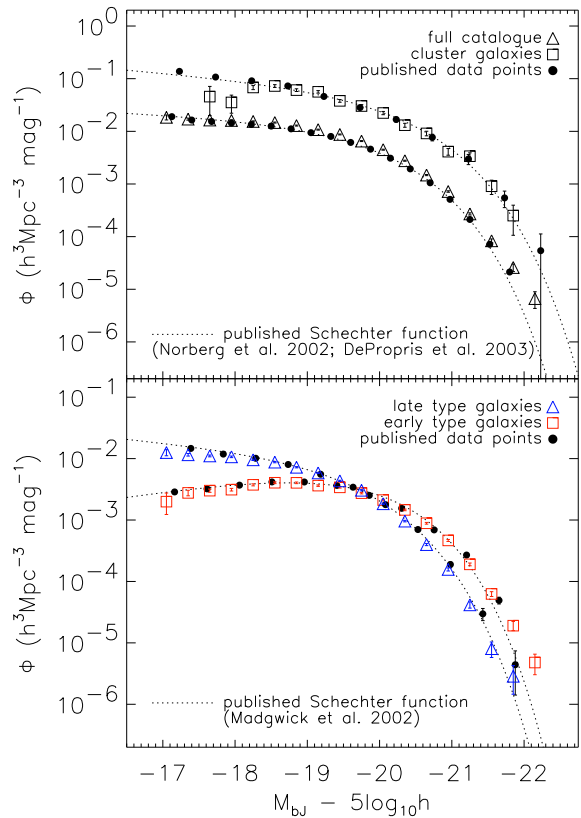}
\caption{A comparison of the published 2dFGRS luminosity
  function (circles and dotted lines) to that calculated by our joint
  SWML(shape)/CiC(normalisation) method (squares and
  triangles) for select galaxy samples.  Shown are the (top) full
  catalogue luminosity function (Norberg et~al.\ 2002a) and cluster
  galaxy population luminosity function (De Propris et~al.\ 2003), and
  (bottom) the luminosity function for early and late-type galaxy
  sub-samples separately (Madgwick et~al.\ 2002).}
\end{figure}

\subsection{Measuring the luminosity function}
The luminosity function, giving the number density of galaxies as a
function of luminosity, is conveniently approximated by the Schechter
function (Schechter 1976, see also Norberg et~al.\ 2002a):
\begin{equation}\label{lfL}
d\phi = \phi^* {(L/L^*)}^{\alpha}\exp(-L/L^*)\ d(L/L^*)~,
\end{equation}
dependent on three parameters: $L^*$ (or equivalently $M^*$),
providing a characteristic luminosity (magnitude) for the galaxy
population; $\alpha$, governing the faint-end slope of the luminosity
function; and $\phi^*$, giving the overall normalisation.  Our method,
which we describe below, will be to use the magnitude-limited
catalogue binned by density and type to calculate the shape of each
luminosity function, draw on restricted volume-limited sub-samples of
each to fix the correct luminosity function normalisation, then
determine the maximum likelihood Schechter function parameters for
each in order to quantify the changing behaviour between different
environments.

\begin{figure*}
\plotfull{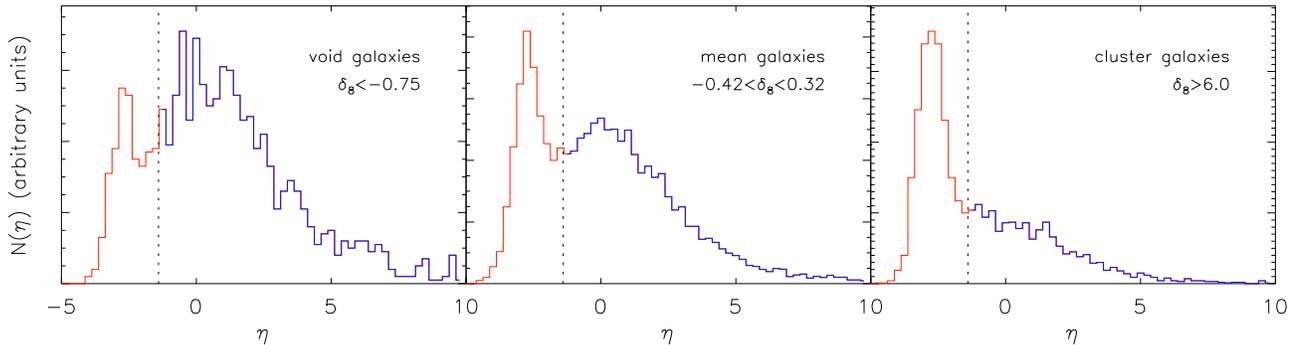}
\caption{The distribution of the spectral parameter, $\eta$, for the
  void, mean, and cluster galaxies used in our 
  analysis (Table~1).  The vertical dotted line at $\eta=-1.4$ divides
  the quiescent galaxy population (early types on left) from the more
  active star-forming galaxies (late types on right).  
  From void to cluster environment, the dominant galaxy
  population changes smoothly from late-type to early-type.
}
\end{figure*}

The luminosity function shape is determined in the standard way using
the step-wise maximum likelihood method (SWML Efstathiou, Ellis \&
Peterson 1988) and the STY estimator (Sandage, Tammann \& Yahil
1979). See Norberg et al. (2002a) for a complete description of the
application of these two techniques to the 2dFGRS. All STY fits are
performed over the magnitude range $-17 > M_{\rm b_J}-5\log_{10} h > -22$.

Such techniques fail to provide the luminosity function normalisation,
however, and one needs to consider carefully how to do this when
studying galaxy populations in different density environments. To
normalise each luminosity function we employ a new counts in cells
(CiC) technique which directly calculates the number density of
galaxies as a function of galaxy magnitude from the galaxy
distribution.  Briefly, this is achieved by counting galaxies in
restricted volume-limited sub-regions of the survey.  We discuss our
CiC method in more detail in Appendix~A.  As we show there, when
galaxy numbers allow a good statistical measurement the luminosity
function shape determined by the SWML and CiC methods agree
very well.
As the SWML estimator draws from the larger magnitude-limited survey
rather than the smaller CiC volume-limited sub-samples, we choose the
above two-step SWML/CiC approach rather than the CiC method alone to
obtain the best results for each luminosity function.  Once the CiC
luminosity function has been calculated for the same galaxy sample,
the SWML luminosity function is then given the correct amplitude by
requiring that the number density integrated between the magnitude
range $-19 > M_{\rm b_J}-5\log_{10} h > -22$ be the same as that for the
CiC result.

\subsection{Comparison to previous 2dFGRS results}

In Fig.~1 we give a comparison of our measured luminosity functions
for selected galaxy populations with the equivalent
previously-published 2dFGRS results (see each reference for complete
details). These include (top panel) the full survey volume (Norberg
et~al.\ 2002a) and cluster galaxy luminosity functions (De Propris
et~al.\ 2003), and (bottom panel) the luminosity functions derived for
late and early-type galaxy populations separately (Madgwick et~al.\
2002).  For all, the squares and triangle symbols show our hybrid
SWML/CiC values while the circles and dotted lines give the
corresponding published 2dFGRS luminosity function data points and
best Schechter function estimates, respectively. The close match between
each set of points confirms that our method is able to reproduce the
published 2dFGRS luminosity shape and amplitude successfully.

There are a few points to note. Firstly, the cluster luminosity
function is not typically quoted with a value of $\phi^*$ since the
normalisation of the cluster galaxy luminosity distribution will vary
from cluster to cluster (dependent on cluster richness).  Because of
this we plot the De Propris et~al.\ cluster luminosity function using
our $\phi^*$ value.

Secondly, the Madgwick et~al.\ early and late-type galaxy absolute
magnitudes include no correction for galaxy evolution, which, if
included, would have the effect of dimming the galaxy population
somewhat. We have checked the significance of omitting the evolution
correction when determining the galaxy absolute magnitudes and
typically find only minimal differences in our results and no change
to our conclusions.

Thirdly, the STY Schechter function values we measure tend to present
a slightly `flatter' faint-end slope than seen for the full survey:
our all-type STY estimate returns $\alpha = -1.05 \pm 0.02$ (Table~1)
whereas for the completed survey (across the redshift range
$0.02<z<0.25$) the recovered $\alpha$ value is $-1.18 \pm 0.02$ (Cole
et al. in preparation).  This difference is due primarily to three
systematic causes: the minimum redshift cut required to define the DDP
which results in a restricted absolute magnitude range over which
we can measure galaxies; the non-perfect description of the galaxy
luminosity function by a Schechter function together with the
existing degeneracies in the $M^*-\alpha$ plane; the sensitivity of
the faint-end slope parametrisation to model dependent corrections for
missed galaxies.
For our results, these systematic effects do not hinder a
comparison between sub-samples, but it is essential to take into
account the different cuts we imposed for any detailed comparison with
other works.
In Appendix~C we discuss these degeneracies and correlations further.
We test their influence by fixing each $\alpha$ at the published field
value when applying the STY estimator and find a typical variation of
less than $0.2$ magnitudes in $M^*$ from the main results presented in
Section~3.  Such systematics do not change our conclusions.

Lastly, the 2dFGRS photometric calibrations have improved since
earlier luminosity function determinations, and thus the good match
seen in Fig.~1 demonstrates that the new calibrations have not
significantly altered the earlier results.

In Fig.~2 we show the $\eta$ distribution for our void, mean, and
cluster galaxy samples.  The mean galaxy distribution is essentially
identical to that shown in Fig.~4 of Madgwick et al. (2002) for the
full survey, demonstrating that the mean density regions contain a
similar mix of galaxy types to that of the survey as a whole.  For
under-dense regions late types progressively dominate, while the
converse is true in the over-dense regions.  This behaviour can be
understood in terms of the density-morphology relation (e.g. Dressler
1980), and will be explored in more detail in the next section.

\begin{figure*}
\plotfull{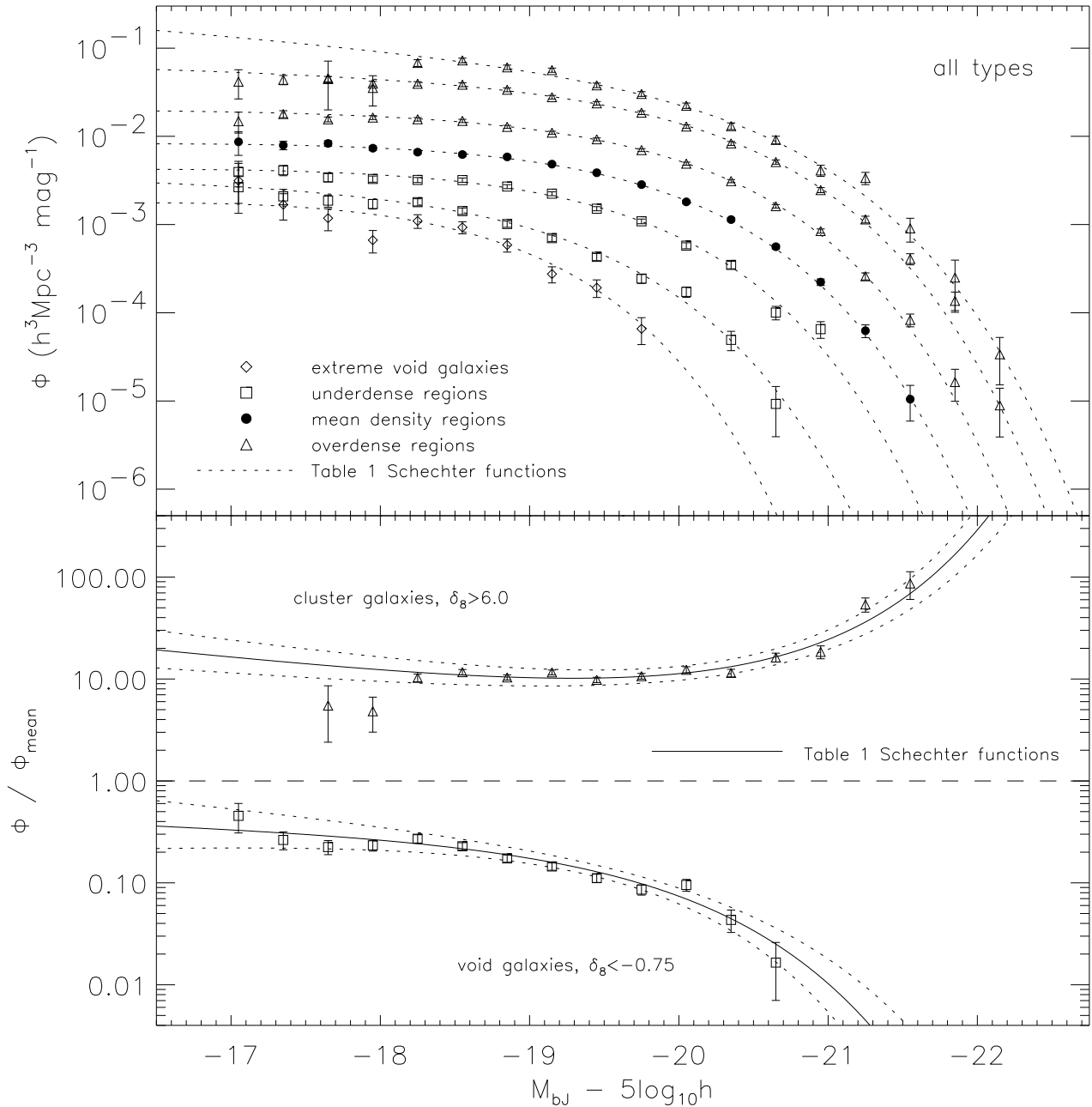}
\caption{(top)~The SWML luminosity functions for the 2dFGRS galaxy
  catalogue in regions of the survey of varying density contrast,
  $\delta_8$, from void to mean density to cluster. The best-fit
  Schechter function parameters for each are given in Table~1 and the
  corresponding Schechter function curves are over plotted here with dotted
  lines. 
  (bottom)~The void and cluster luminosity functions 
  normalised to the mean luminosity function so as to highlight the relative
  differences in the shape of each distribution.  The solid lines and
  bounding dotted lines show the appropriate Table~1 Schechter
  functions normalised to the mean Schechter function and $1 \sigma$
  uncertainty.
}
\end{figure*}

\begin{table*}
\centering
\footnotesize
\caption{
  Properties of our magnitude-limited galaxy samples, split by
  spectral type (all, early and late) and in seven density ranges
  (defined by $\delta_{8_{\rm{min}}}$ and $\delta_{8_{\rm{max}}}$, the
  density contrast in spheres of radius $8h^{-1}$Mpc). The all-type
  sample is also split into an `extreme void' sample. $\rm{N_{GAL}}$
  and $f_{\rm{VOL}}$ are the number of galaxies in each density bin
  and the volume fraction these galaxies occupy,
  respectively. $f_{\rm{VOL}}$ is given for all galaxy types only:
  early/late-type density populations are just sub-divisions of the
  associated all-type sample. $M^*$ and $\alpha$ are the likelihood
  estimated Schechter function parameters, and $\phi^*$ the associated
  normalisation. The integrated luminosity density, as defined by
  Eq.~3 with $L_{min} = 0$, is given in the last column. All errors on
  the derived parameters reflect only the associated statistical
  uncertainty.
}
\begin{scriptsize}
\begin{tabular}{llccrccccc} 
\hline \hline
 Galaxy & Galaxy & $\delta_{8_{\rm min}}$ & $\delta_{8_{\rm max}}$ & {N$_{\rm GAL}$} & {$f_{\rm VOL}$} & \hspace{-0.2cm}{$M^*$} &\hspace{-0.2cm} {$\alpha$} &\hspace{-0.2cm}{$\phi^*$} &\hspace{-0.2cm}\smash{$\langle\rho_L\rangle$} \\
 Type   & Sample &   &   &   &   &   {\tiny $M_{b_{\rm J}}-5\log_{10}h$} &     & {\tiny $10^{-3}h^{3}$Mpc$^{-3}$} & {\tiny $10^{8}h$L$_{\odot}$Mpc$^{-3}$}      \\
\hline \hline
   all types: & full volume & \hspace{-0.2cm}$-1.0$  & \hspace{-0.0cm}$\infty$  & $81,387$ & $1.0$  & \hspace{-0.2cm}$-19.65\pm0.02$ & \hspace{-0.2cm}$-1.05\pm0.02$ & \hspace{-0.33cm} $21.3\pm0.5$  & \hspace{-0.2cm}  $2.10\pm0.08$  \\
 &             extreme void & \hspace{-0.2cm}$-1.0$  & \hspace{-0.2cm}$-0.90$   & $260$     & $0.09$ & \hspace{-0.2cm}$-18.26\pm0.33$ & \hspace{-0.2cm}$-0.81\pm0.50$ & \hspace{-0.2cm}  $3.17\pm0.90$ & \hspace{-0.2cm}  $0.08\pm0.04$  \\
\hline                                                                                                                                                                                                                             
 &                     void & \hspace{-0.2cm}$-1.0$  & \hspace{-0.2cm}$-0.75$   & $1,157$   & $0.20$ & \hspace{-0.2cm}$-18.84\pm0.16$ & \hspace{-0.2cm}$-1.06\pm0.24$ & \hspace{-0.2cm}  $3.15\pm0.56$ & \hspace{-0.2cm} $0.15\pm0.04$   \\
 &                          & \hspace{-0.2cm}$-0.75$ & \hspace{-0.2cm}$-0.43$   & $3,331$   & $0.19$ & \hspace{-0.2cm}$-19.20\pm0.10$ & \hspace{-0.2cm}$-0.93\pm0.11$ & \hspace{-0.2cm}  $5.99\pm0.62$ & \hspace{-0.2cm} $0.36\pm0.05$   \\
 &                     mean & \hspace{-0.2cm}$-0.43$ & \hspace{-0.0cm}$0.32$    & $11,877$  & $0.30$ & \hspace{-0.2cm}$-19.44\pm0.05$ & \hspace{-0.2cm}$-0.94\pm0.05$ & \hspace{-0.33cm} $11.3\pm0.7$  & \hspace{-0.2cm} $0.86\pm0.07$   \\
 &                          & \hspace{-0.0cm}$0.32$  & \hspace{-0.0cm}$2.1$     & $21,989$  & $0.24$ & \hspace{-0.2cm}$-19.64\pm0.04$ & \hspace{-0.2cm}$-0.99\pm0.04$ & \hspace{-0.33cm} $22.9\pm1.0$  & \hspace{-0.2cm} $2.16\pm0.13$   \\
 &                          & \hspace{-0.0cm}$2.1$   & \hspace{-0.0cm}$6.0$     & $15,656$  & $0.07$ & \hspace{-0.2cm}$-19.85\pm0.05$ & \hspace{-0.2cm}$-1.09\pm0.04$ & \hspace{-0.33cm} $49.0\pm3.0$  & \hspace{-0.2cm} $5.95\pm0.49$   \\
 &                  cluster & \hspace{-0.0cm}$6.0$   & \hspace{-0.0cm}$\infty$  & $3,175$   & $0.01$& \hspace{-0.2cm}$-20.08\pm0.13$ & \hspace{-0.2cm}$-1.33\pm0.11$ & \hspace{-0.1cm}$60.7\pm13.2$   & \hspace{-0.22cm}$11.6\pm3.4$    \\
\hline                                                                                                                                                                              
   late type: & full volume & \hspace{-0.2cm}$-1.0$  & \hspace{-0.0cm}$\infty$  & $42,772$  & $  - $  & \hspace{-0.2cm}$-19.30\pm0.03$ & \hspace{-0.2cm}$-1.03\pm0.03$ & \hspace{-0.33cm}  $15.0\pm0.5$ & \hspace{-0.2cm} $1.06\pm0.05$   \\
\hline                                                                                                                                                                                                          
 &                     void & \hspace{-0.2cm}$-1.0$  & \hspace{-0.2cm}$-0.75$   & $855$   & $  - $ & \hspace{-0.2cm}$-18.78\pm0.19$ & \hspace{-0.2cm}$-1.14\pm0.24$ & \hspace{-0.2cm}  $2.42\pm0.55$ & \hspace{-0.2cm} $0.11\pm0.04$   \\
 &                          & \hspace{-0.2cm}$-0.75$ & \hspace{-0.2cm}$-0.43$   & $2,249$   & $  - $ & \hspace{-0.2cm}$-19.07\pm0.12$ & \hspace{-0.2cm}$-0.95\pm0.14$ & \hspace{-0.2cm}  $4.54\pm0.58$ & \hspace{-0.2cm} $0.25\pm0.05$   \\
 &                     mean & \hspace{-0.2cm}$-0.43$ & \hspace{-0.0cm}$0.32$    & $7,261$   & $  - $ & \hspace{-0.2cm}$-19.24\pm0.07$ & \hspace{-0.2cm}$-1.00\pm0.07$ & \hspace{-0.2cm}  $8.03\pm0.61$ & \hspace{-0.2cm} $0.53\pm0.06$   \\
 &                          & \hspace{-0.0cm}$0.32$  & \hspace{-0.0cm}$2.1$     & $11,921$  & $  - $ & \hspace{-0.2cm}$-19.36\pm0.06$ & \hspace{-0.2cm}$-1.04\pm0.05$ & \hspace{-0.33cm}  $15.5\pm1.0$ & \hspace{-0.2cm} $1.17\pm0.11$   \\
 &                          & \hspace{-0.0cm}$2.1$   & \hspace{-0.0cm}$6.0$     & $7,596$   & $  - $ & \hspace{-0.2cm}$-19.37\pm0.07$ & \hspace{-0.2cm}$-1.03\pm0.07$ & \hspace{-0.33cm}  $36.3\pm2.9$ & \hspace{-0.2cm} $2.73\pm0.31$   \\
 &                  cluster & \hspace{-0.0cm}$6.0$   & \hspace{-0.0cm}$\infty$  & $1,316$   & $  - $ & \hspace{-0.2cm}$-19.34\pm0.18$ & \hspace{-0.2cm}$-1.09\pm0.20$ & \hspace{-0.2cm}  $54.0\pm12.2$ & \hspace{-0.2cm} $4.09\pm1.31$   \\
\hline                                                                                                                                                                              
  early type: & full volume & \hspace{-0.2cm}$-1.0$  & \hspace{-0.0cm}$\infty$  & $30,354$  & $  - $ & \hspace{-0.2cm}$-19.65\pm0.03$ & \hspace{-0.2cm}$-0.65\pm0.03$ & \hspace{-0.2cm} $8.80\pm0.22$  & \hspace{-0.2cm} $0.75\pm0.03$   \\
\hline                                                                                               
 &                     void & \hspace{-0.2cm}$-1.0$  & \hspace{-0.2cm}$-0.75$   & $220$     & $  - $ & \hspace{-0.2cm}$-18.62\pm0.33$ & \hspace{-0.2cm}$-0.15\pm0.53$ & \hspace{-0.2cm} $0.67\pm0.10$ & \hspace{-0.2cm} $0.02\pm0.01$    \\
 &                          & \hspace{-0.2cm}$-0.75$ & \hspace{-0.2cm}$-0.43$   & $861$   & $  - $ & \hspace{-0.2cm}$-19.16\pm0.14$ & \hspace{-0.2cm}$-0.43\pm0.24$ & \hspace{-0.2cm} $1.62\pm0.17$ & \hspace{-0.2cm} $0.88\pm0.02$    \\
 &                     mean & \hspace{-0.2cm}$-0.43$ & \hspace{-0.0cm}$0.32$    & $3,873$   & $  - $ & \hspace{-0.2cm}$-19.38\pm0.08$ & \hspace{-0.2cm}$-0.39\pm0.11$ & \hspace{-0.2cm} $4.13\pm0.19$ & \hspace{-0.2cm} $0.27\pm0.02$    \\
 &                          & \hspace{-0.0cm}$0.32$  & \hspace{-0.0cm}$2.1$     & $8,809$  & $  - $ & \hspace{-0.2cm}$-19.59\pm0.05$ & \hspace{-0.2cm}$-0.52\pm0.06$ & \hspace{-0.33cm} $10.6\pm0.4$ & \hspace{-0.2cm} $0.84\pm0.05$    \\
 &                          & \hspace{-0.0cm}$2.1$   & \hspace{-0.0cm}$6.0$     & $7,163$   & $  - $ & \hspace{-0.2cm}$-19.89\pm0.06$ & \hspace{-0.2cm}$-0.81\pm0.06$ & \hspace{-0.33cm} $24.2\pm1.6$ & \hspace{-0.2cm} $2.67\pm0.23$    \\
 &                  cluster & \hspace{-0.0cm}$6.0$   & \hspace{-0.0cm}$\infty$  & $1,731$   & $  - $ & \hspace{-0.2cm}$-20.13\pm0.18$ & \hspace{-0.2cm}$-1.12\pm0.14$ & \hspace{-0.33cm} $37.1\pm7.7$ & \hspace{-0.2cm} $6.00\pm1.75$    \\
\hline \hline
\end{tabular}
\end{scriptsize}
\end{table*}

\section{RESULTS}

\subsection{Luminosity functions}

The top panel of Fig.~3 shows the 2dFGRS galaxy luminosity function
estimated for the six logarithmically-spaced density bins and
additional extreme void bin, $\delta_8 < -0.9$, given in Table~1.  The
luminosity function varies smoothly as one moves between the extremes
in environment. Each curve shows the characteristic shape of the
Schechter function, for which we show the STY fit across the entire
range of points plotted with dotted lines.  The Schechter parameters
are given in Table~1, along with the number of galaxies considered in
each density environment and the volume fraction they occupy. A number
of points of interest regarding the variation of these parameters with
local density will be discussed below.

To examine the relative differences in the void and cluster galaxy
populations with respect to the mean, in the bottom panel of Fig.~3 we
plot the ratio of the void and cluster luminosity functions to the
mean luminosity function.  Also shown is the ratio of the corresponding
Schechter functions to the mean Schechter function (solid lines) and
$1\sigma$ uncertainty (dotted lines, where only the error in $M^*$ and
$\alpha$ has been propagated).
For a non-changing luminosity function shape, this ratio is a
flat line whose amplitude reflects the relative abundance of the
samples considered. For two Schechter functions differing in both
$\alpha$ and $M^*$, the faint-end of the ratio is most sensitive to the
differences in $\alpha$ and the bright end to the
differences in $M^*$. 
We note that the error regions on the Schechter function fits shown
here do not include the uncertainty of the mean sample, as the
correlation of its error with the other samples is unknown.  This
panel reveals significant shifts in abundances at the bright end: in
voids there is an increasing deficit of bright galaxies for magnitudes
$M_{\rm b_J} -5 \log h \la - 18.5$, while clusters exhibit an excess
of very luminous galaxies at magnitudes $M_{\rm b_J} -5 \log h \la -
21$.

It is well-established that early and late-type galaxy populations
have very different luminosity distributions (Fig.~1).
In Fig.~4 we explore the density dependence of these populations.  The
upper panels show the luminosity functions and their Schechter
function fits, as in Fig.~3, but for (left panel) early types and
(right panel) late types separately.  In the corresponding lower
panels we show the ratio of each extreme density population to the
mean density luminosity function, following the same format as the
bottom panel of Fig.~3.  (We note that the mean luminosity functions
for each type used in this figure are both very similar in shape to
that shown in the bottom panel of Fig.~1).  The best-fit Schechter
parameters are given in Table~1.  Again we see a smooth transition in
the galaxy luminosity function as one moves through regions of
different density contrast.  The lower left panel of Fig.~4 shows a
significant variation of the bright end early-type galaxy population
with respect to the mean, while at the faint end the changes are more
ambiguous, but with Schechter fits that suggest some
evolution into the denser regions.  Note that, although the faint end
of our early-type cluster Schechter function is primarily constrained
by the mid-luminosity galaxies in the sample, our maximum liklihood
Schechter parameters are quite close to that found by De Propris et
al. (2003) for a comparable galaxy population but measured
approximately three magnitudes fainter.  
In contrast to the early types, in the lower right panel of Fig.~4
late-type galaxies show little change in relative population between
the mean and cluster environments and a possible ``tilt" favouring the
faint-end for low-density environments.  Due to deteriorating
statistics we do not consider the type dependent extreme void
luminosity function which was introduced in Fig.~3.

The essence of our results is best appreciated when we directly
compare the early and late-type galaxy distributions, separately for
the cluster and void regions of the survey, as shown in Fig.~5. This
figure reveals a striking contrast: the void population is composed
primarily of medium to faint luminosity late-type galaxies, while for
the cluster population early types dominate down to all but the
faintest magnitude considered.  This is the central result of our
study, 
and shows the crucial role of accurately determining the
amplitude of the luminosity function, since the shape alone does not
necessarily determine the dominant population of a region.

\begin{figure*}
\plotfull{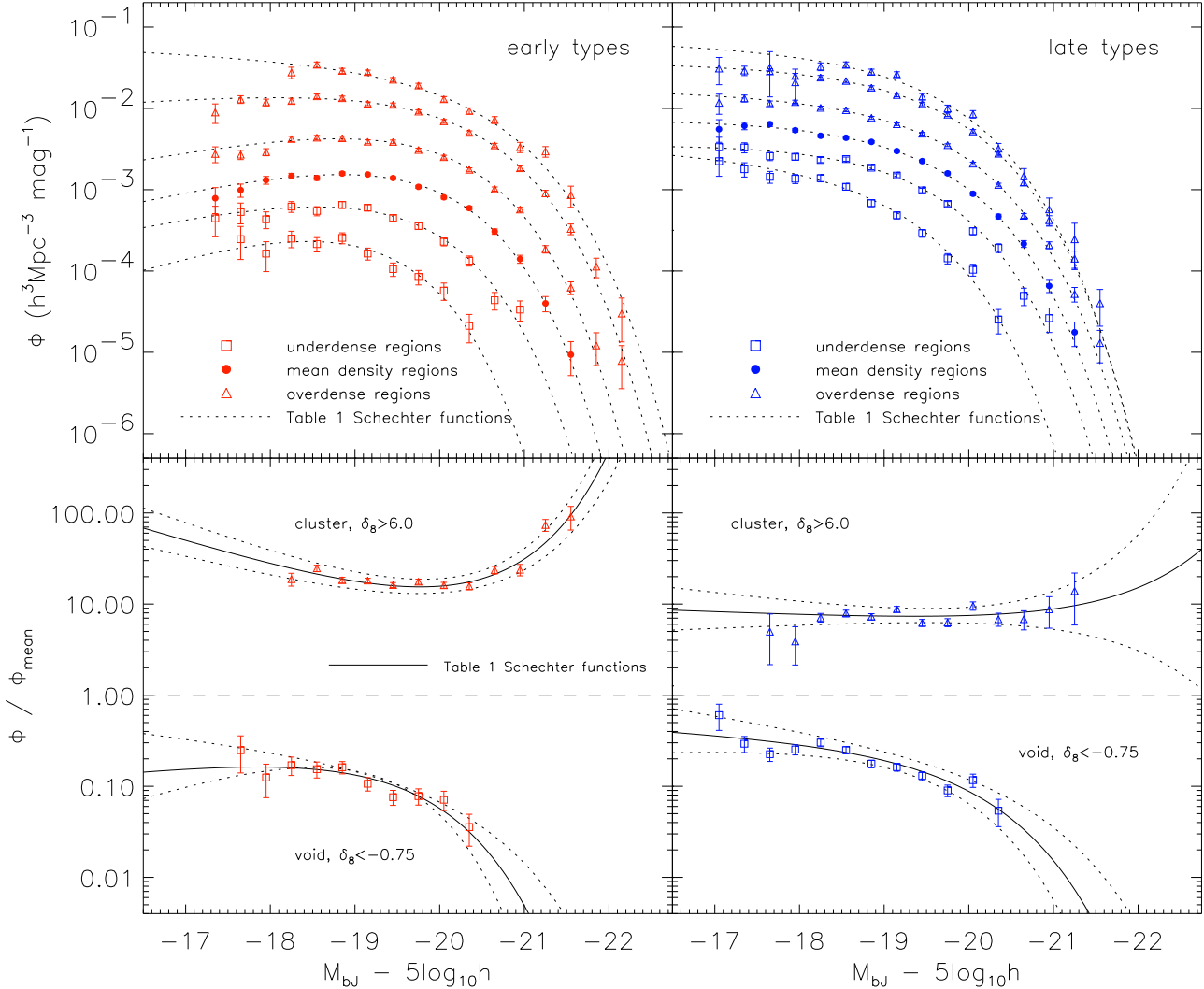}
\caption{Comparing both the (top) absolute and (bottom) relative
  distributions of (left) early-type galaxies in different density
  environments, and (right) late-type galaxies in different density
  environments.  In the bottom panels the luminosity functions have
  again been normalised to the mean (each to their respective type) as
  done previously in Fig.~3 (note that the shape of the mean for each
  type is very similar to that shown in Fig.~1).  Here the solid lines
  and bounding dotted lines show the appropriate Table~1 Schechter
  functions normalised to the mean Schechter function and $1 \sigma$
  uncertainty.
}
\end{figure*}

\begin{figure}
\plotone{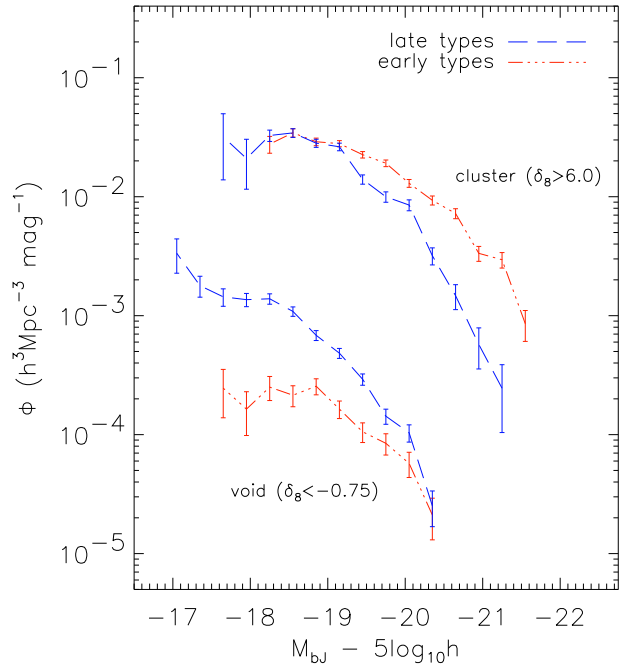}
\caption{A direct comparison of the early and late-type galaxy
  populations in the cluster environment (top two luminosity
  functions) and void regions of the survey (bottom two luminosity
  functions). The void population is composed almost exclusively of
  faint late-type galaxies, while in the clusters regions the galaxy
  population brighter than $M_{\rm b_J}-5\log_{10} h = -19$ consists
  predominantly of early types.}  
\end{figure}

\subsection{Evolution with environment}

It is well known that the Schechter function parameters are highly
correlated.  In Fig.~6 we show the $1\sigma$ ($68\%$ $2$-parameter)
and $3\sigma$ ($99\%$ $2$-parameter) $\chi^2$ contours in the $M^* -
\alpha$ plane for the early-type, late-type, and combined type cluster
and void populations.  For a given spectral type, all show a greater
than $3\sigma$ difference in the STY Schechter parameters between the void
and cluster regions.  Intermediate density bins are omitted for
clarity but follow a smooth progression with smaller error ellipses
between the two extremes shown.  In Appendix~C we explore in more
detail the $M^* - \alpha$ degeneracy and confirm that our results are
robust.

Our findings show that the galaxy luminosity function changes
gradually with environment.  We quantify this behaviour in Fig.~7 by
plotting the variation of $M^*$ and $\alpha$ as a function of density
contrast, where points to the left of $\delta_8 = 0$ represent the
under-dense to void regions in the survey, and points to the right of
this are measured in the over-dense to cluster regions.  Late-type
galaxies display a consistent luminosity function across all density
environments, from sparse voids to dense clusters, with a weak dimming
of $M^*$ in the under-dense regions, and an almost constant faint-end
slope. In contrast, the luminosity distribution of early-type galaxies
differs sharply between the extremes in environment: $M^*$ brightens
by approximately $1.5$ magnitudes going from voids to clusters, while
the faint-end slope moves from $\alpha \approx -0.3$ in under-dense
regions to around $\alpha\approx-1.0$ in the densest parts of the
survey.

Finally, in Fig.~8 we plot the mean luminosity per galaxy,
$\langle\rho_L\rangle / \langle\rho_g\rangle$, obtained by integrating
the luminosity function for each set of Schechter function parameters
from Table~1:
\begin{equation}\label{rhog}
\langle\rho_g\rangle = \int^{\infty}_{L_{min}} \phi(L)\ dL~,\ \ \ 
\langle\rho_L\rangle = \int^{\infty}_{L_{min}} \phi(L) L\ dL~,
\end{equation}
where $L_{min}$ is a somewhat arbitrary observational cutoff chosen at
$M_{\rm b_J} -5\log_{10}h = -17$. This is both the limit down to which we
confidently measure our luminosity functions, and also the limit
beyond which the Schechter function no longer provides a good fit to
the early-type luminosity function of Madgwick et~al. (2002). The
final column of Table~1 gives the total luminosity density in the
various density contrast environments, computed by integrating the
Schechter function with no cutoff, to allow easy comparison with past
and future analyses; the contribution to the calculated
$\langle\rho_L\rangle$ from luminosities below the observational
cutoff is less than a few percent. We note that $\langle\rho_g\rangle$ is
directly related to the density contrast, $\delta_8$, by definition.
It is interesting to see that the early-type galaxies in Fig.~8 are,
on average, about a factor of two brighter per galaxy than the late
types, even though the late types dominate in terms of both number and
luminosity density.  For all galaxy populations, the mean luminosity
per galaxy shows a remarkable constancy across the full range of
density environments.

\section{COMPARISON TO PREVIOUS WORK}

\begin{figure}
\plotone{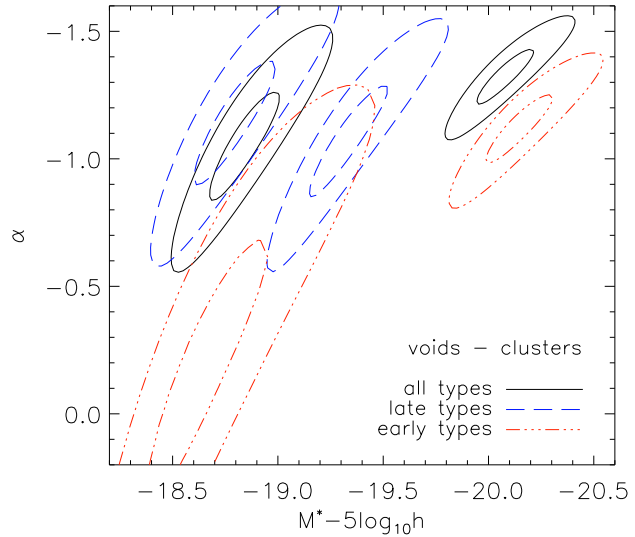}
\caption{The $1\sigma$ ($68\%$ $2$-parameter) and $3\sigma$ ($99\%$
  $2$-parameter) contours of constant $\chi^2$ in the $M^* - \alpha$
  plane for the void and cluster STY estimates (in each case the void fit
  is on the left, corresponding to a fainter M*).  Galaxy types are
  identified by the linestyle given in the legend. Even at the
  $3\sigma$ level significant differences in the void and cluster
  Schechter function parameters for each galaxy type can be seen. }
\end{figure}

Historically, work on the dependence of the luminosity function on
large scale environment has been restricted primarily to comparisons
between cluster and field galaxies, due to insufficient statistics to
study voids. (Note that `field' samples are usually flux-limited
catalogues which cover all types of environments.) One of the aims of this
work is to elucidate the properties of galaxies in void environments and
understand the relationship between 
clusters and voids. 
In this section, we briefly summarise previous observations and
compare them with the results presented in Section~3.

We have already shown in Fig.~1 and Section~2 that our cluster and
field results are equivalent to the published 2dFGRS luminosity
function results of Norberg et~al.  (2002a), Madgwick et~al.\ (2002),
and De Propris et~al.\ (2003). The latter paper explains their cluster
luminosity function by demonstrating that the field luminosity
function can be approximately transformed into the cluster luminosity
function using a simple model where the cluster environment suppresses
star formation to produce a dominant bright, early-type population
(see Section 4.4 of their paper for details).  We expand upon such
models in the next section.

Bromley et~al.\ (1998) considered $18,278$ galaxies in the Las Campanas
Redshift Survey (LCRS) as a function of spectral type and high and low
local density. We confirm (e.g., Fig.~7) their qualitative finding
that for early-type galaxies the faint-end slope steepens with density
whereas late-type objects show little or no significant trend. We cannot
make a quantitative comparison to their result because they do not give
the definition of their low density sample. 

H\"{u}tsi et~al. (2003) use the Early Data Release of the Sloan
Digital Sky Survey (SDSS) and the LCRS to consider the galaxy
luminosity function as a function of density field, but in
two-dimensional projection so their results are not directly
comparable to ours.  They find a faint-end slope of $\alpha \approx
-1.1$ in all environments and an increase in $M^*$ of roughly $0.3$
magnitudes between the under and over-dense portions of their data.
This is broadly consistent with the more detailed results obtained
here with the full 2dFGRS catalogue when one averages over our two
most underdense bins and three most over-dense.  In separate work,
these authors also consider the environmental dependence of cluster
and supercluster properties in the SDSS and LCRS (Einasto et
al. 2003a, 2003b).  They show an almost order of magnitude increase in
the mean cluster luminosity between extremes in density (defined in
two dimensions by smoothing over a projected $10h^{-1}$Mpc radius
around each cluster).  A comparison of their results to ours
(i.e. Fig.~7) suggests a correlation between galaxy, galaxy group, and
galaxy cluster properties in a given density environment.  A more
detailed exploration would shed light on the connection between
virialised objects of different mass with local density.  We defer
such an investigation to later work.

\begin{figure}
\plotone{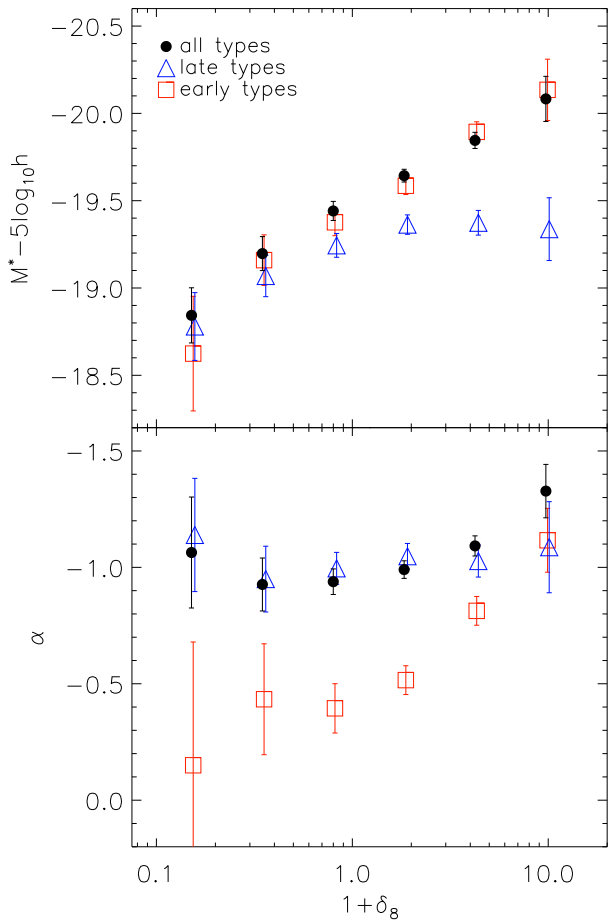}
\caption{The maximum liklihood Schechter function $M^*$ and $\alpha$ parameters
  for each of the six density contrast regions in Table~1 (Figs.~3 and 4). 
  Each panel shows the result for individual samples split by spectral type
  (early/late) and both types combined.
}
\end{figure}

\begin{figure}
\plotone{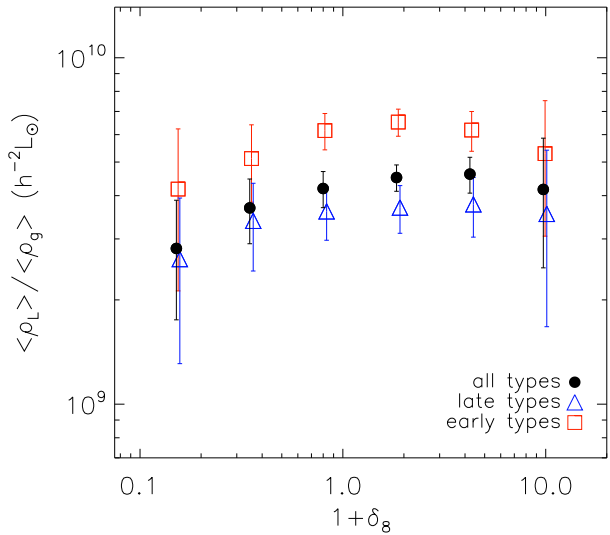}
\caption{The mean luminosity per galaxy as a function of 
  density environment for each galaxy type, calculated from
  Eq.~\ref{rhog} using the Schechter function parameters given in
  Table~1. The integral is performed by choosing $L_{min}$ so that
  $M_{min}-5\log_{10}h=-17$ (Section~3.2). The mean luminosity per
  galaxy of early types is consistently about a factor of two brighter
  than their late type counterparts across all density environments.}
\end{figure}

In a series of papers, members of the SDSS team undertook an analysis
of the properties of galaxy samples drawn from under-dense regions in
the SDSS (Rojas et al. 2004, 2003; Goldberg et al. 2004; Hoyle
et al. 2003).  Of most relevance to our study is the work of
Hoyle et al. who completed a preliminary analysis of the SDSS void
luminosity function, defined in regions of $\delta_7 < -0.6$ using a
smoothing scale of $7h^{-1}$Mpc.  Their sample of $1,010$ void
galaxies are typically fainter and bluer than galaxies in higher
density environments but with a similar faint-end slope.  Their
results are consistent with what we find using a sample which contains
about twice the number of void galaxies as defined by Hoyle et al..
Using the same void galaxy catalogue, Rojas et al. (2003) show
that this behaviour is not merely an extrapolation of the
density-morphology relationship (e.g. Dressler 1980) into sparser
environments.  By measuring the concentration and Sersic indices
(Sersic 1968, Blanton et al. 2003b) of void and field galaxies they
detect no significant shift in the morphological mix, even though
their void galaxy sample is shown to be significantly bluer.

Also using the SDSS dataset, Hogg et al. (2003) consider the mean
environment as a function of luminosity and colour of $115,000$
galaxies, on smoothing scales of $1$ and $8h^{-1}$Mpc.  They find that
their reddest galaxies show strong correlations of luminosity with
local density at both the faint and bright extremes, whereas the
luminosities of blue galaxies have little dependence on environment.
These conclusions are consistent with the present results for our
early (red) and late (blue) type luminosity functions (Fig.~7).
However by restricting attention to the {\it average} environment of a
galaxy of given luminosity and colour, their sample is by definition
dominated by galaxies in over-dense environments.  The measures they
consider are therefore insensitive to one of the main questions of
interest to us here, namely whether the characteristic galaxy
population in the voids is distinctively different from that in other
density environments.  Indeed, we clearly find evidence for a
population which is particularly favoured in void regions, namely
faint late-type galaxies (Fig.~5).

\section{DISCUSSION}

\begin{table*}
\centering
\footnotesize
\caption{A mnemonic summary of our main results, drawing on
  the work of De Propris et~al.\ (2003) and Mo et~al.\ (2004) to
  interpret the observed behaviour in Figs.~5 and 7 in terms of
  physical processes which govern the void and cluster galaxy
  populations.}
\begin{tabular}{ccl} 
\hline \hline
Region   & Observation            & Process                                                      \\
\hline \hline    
Voids    &                        & 1.~galaxies typically reside at the centre                   \\
         & faint, late type       & ~~~of low mass dark halos ($\Rightarrow$ faint)              \\
         & galaxies dominate      & 2.~gas is available for star formation ($\Rightarrow$ blue)  \\
         &                        & 3.~merger rate is low ($\Rightarrow$ spirals)                \\
\hline                                                                                           
Clusters &                        & 1.~typically satellite and central galaxies of               \\
         & mid-bright, early-type & ~~~massive dark halos ($\Rightarrow$ mid-bright)             \\
         & galaxies dominate      & 2.~gas is unavailable for star formation ($\Rightarrow$ red) \\
         &                        & 3.~merger rate is high ($\Rightarrow$ ellipticals)           \\
\hline \hline
\end{tabular}
\end{table*}  

As clusters are comparatively well-studied objects, and have 
already been addressed using the 2dFGRS by De Propris et al. (2003), we 
focus here primarily on a discussion of the voids.

A detailed analysis of void population properties has recently become
possible due to significant improvements in the quality of both
theoretical modelling and observational data, as summarised by Benson
et~al.\ (2003). Peebles (2001) has argued that, visually, observed
voids do not match simulated ones and discussed several statistical
measures for quantifying a comparison, primarily the distance to the
nearest neighbour in a reference sample. However the cumulative
distributions of nearest neighbour distances shown in Figs.~4--6 of
Peebles (2001) show very little difference between the
reference--reference and test--reference distributions. It is not
surprising that these statistical measures are insensitive to a void
effect, since they are dominated by cluster galaxies. Our method is
designed to overcome this difficulty by explicitly isolating the void
population of galaxies so that their properties can be studied.

Motivated by the claims of discrepancies in Peebles (2001), Mathis \&
White (2002) investigated the nature of void galaxies using N-body
simulations with semi-analytic recipes for galaxy formation. 
They call into question the assertion of Peebles (2001) that
$\Lambda$CDM predicts a population of small haloes in the voids,
concluding that ``the population of faint galaxies...does not
constitute a void population''.  More specifically, they find that all
types of galaxies tend to avoid the void regions of their simulation,
down to their resolution limit of $M_{\rm B}= -16.27$ in luminosity
and  $M_{\rm B} = -18.46$ in morphology.

The abundance of faint galaxies we find in the void regions of the
2dFGRS seems to be at odds with the Mathis \& White predictions.
However their results rely on the Peebles (2001) cumulative
distribution of galaxies as a function of over density (their Fig.~3)
which, like cumulative distributions in general, are rather
insensitive to numerically-minor components of the galaxy population.
Note that Mathis \& White define density contrast using the dark
matter mass distribution smoothed over a $5~h^{-1}$Mpc sphere, whereas
we measure the density contrast by galaxy counts.  Another possible
source of discrepancy is the uncertainties in their semi-analytic
recipes, such as the implementation of supernova feedback, which can
strongly effect the faint-end luminosity distribution.  

There has been recent discussion in the literature about the nature of
the faint-end galaxy population and its dependence on group and
cluster richness. Most notably, Tully et al. (2002) show a significant
steepening in the faint-end population as one considers nearby galaxy
groups of increasing richness, from the Local Group to Coma.  On the
surface of it, this might seem at variance with our results, which are
rather better described by a faint-end slope which is approximately
constant with changing density environment for the full 2dFGRS galaxy
sample (Fig.~7).  However the steepening of the faint-end slope they
find primarily occurs at magnitudes fainter than $M_{\rm B} = -17$,
which is beyond the limit we can study with our sample.  Also, their
analysis focuses on individual groups of galaxies, while we have
chosen to work with a much bigger galaxy sample and have smoothed it
over a scale much larger than the typical cluster.  Indeed, as
discussed in Appendix B, when the smoothing scale is significantly larger than
the characteristic size of the structures being probed
it is possible that the Schechter function parameters may become
insensitive to the small scale shifts in population.
This effect, of course, would be less significant for survey regions
which host clusters of clusters (i.e. super-clusters), and which are
prominently seen in the 2dFGRS (Baugh et al. 2004, Croton et
al. 2004).
When sampling the 2dFGRS volume the trend with density that one sees
using $4 h^{-1}$ Mpc spheres in Fig.~B1 is consistent with the Tully
et al. result, although one needs to additionally understand the
influences of Poisson noise.

Tully et al. attribute their results to a process of photoionisation of
the IGM which suppresses dwarf galaxy formation.  Over-dense regions,
which at later times become massive clusters, typically collapse early
and thus have time to form a dwarf galaxy population before the epoch
of reionisation.  Under-dense regions, on the other hand, begin their
collapse at much later times and are thus subject to the
photoionisation suppression of cooling baryons.  This, they argue,
explains the significant increase between the dwarf populations of the
Local Group (low density environment) and Coma (over-dense
environment).  Although suggestive, a deeper understanding of what is
happening will require a much more statistically significant sample.

Recently Mo et~al.\ (2004) have considered the dependence of the
galaxy luminosity function on large scale environment in their halo
occupation model. In this model, the mass of a dark matter halo alone
determines the properties of the galaxies. They create mock catalogues
built with a halo-conditional luminosity function (Yang et al. 2003)
which is constrained to reproduce the overall 2dFGRS luminosity
function and correlation length for both luminosity and type. They
analyse their data by smoothing over spheres of radius $8~h^{-1}$Mpc
in their mock catalogue, and measure the luminosity function as a
function of density contrast. Their work is performed in real space
while we are restricted to work in redshift space.  Nonetheless, their
predictions qualitatively match our density-dependent luminosity
functions; a quantitative comparison is deferred to subsequent work.

In the framework of the Mo et al. model, the reason the faint-end
slope $\alpha$ has such a strong dependence on local density
for early types (Fig.~7) is that faint ellipticals tend to reside
predominantly in cluster-sized halos. The $\alpha$ dependence is
weaker for late types because faint later-type galaxies tend to live
primarily in less massive haloes, which are present in all density
environments.  The correlations between dark halo mass and the
properties of the associated galaxies are not a fundamental prediction
of their model, but are input through phenomenological functions
adjusted to give agreement with the 2dFGRS overall luminosity
functions by galaxy type.  However it would be a non-trivial result if
the correlations are the same independently of whether the dark matter
halo is in a void or in a cluster.  For instance, this property would
not apply in models for which reionisation more efficiently prevented
star formation in under-dense environments than in over-dense
environments, as discussed above.

An interesting consequence of the Mo et~al.\ (2004) model is that the
luminous galaxy distribution (which is easy to observe but hard to
model) correlates well with the dark halo mass distribution (which is
hard to observe but easy to model).
If their predictions prove to give a good description of the present
data it will lend credence to the underlying assumption of their
model---that the environmental dependence of many fundamental galaxy
properties are entirely due to the dependence of the dark halo mass
function on environment.  Exactly why this is would still need to be
explained, however such a demonstration may facilitate more detailed
comparisons between theory and observation than previously possible.

An important result of our work is presented in Fig.~5, where a
significant shift in the dominant population between voids and
clusters is seen.  Such a result points to substantial differences in
the evolutionary tracks of cluster and void early-type galaxies.
Cluster galaxies have been historically well studied: they are more
numerous and much brighter on average, with an evolution dominated by
galaxy-galaxy interactions and mergers.  In voids, however, the
picture is not so clear.  A reasonable expectation would be that the
dynamical evolution of void galaxies should be much slower due to
their relative isolation, with passively evolved stellar populations
and morphologies similar to that obtained during their formation.
Targeted observational studies of void early-type galaxies may reveal
much about the high redshift formation processes that go into making
such rare objects.

Table~2 summarises our main results and provides a qualitative or
mnemonic interpretation based on our observations and the work of Mo
et~al.\ (2004) and De Propris et~al.\ (2003) (and references therein).
Our primary result is the striking change in population types between
voids and clusters shown in Fig.~5: faint, late-type galaxies are
overwhelmingly the dominant galaxy population in voids, completely the
contrary of the situation in clusters.  The existence of such a
population in the voids, and more generally the way populations of
different type are seen to change between different density
environments, place important constraints on current and future models
of galaxy formation.

\section*{ACKNOWLEDGEMENTS}

The 2dFGRS was undertaken using the two-degree field spectrograph on
the Anglo-Australian Telescope. We acknowledge the efforts of all
those responsible for the smooth running of this facility during the
course of the survey, and also the indulgence of the time allocation
committees.  Many thanks go to Simon White, Frank van den Bosch,
Guinevere Kauffmann, and Jim Peebles for useful discussions.  
We also wish to thank the referee, Jaan Einasto, for swift and
constructive comments which improved the content of the paper.
DJC wishes to thank the NYU Center for Cosmology and Particle Physics
for its hospitality during part of this research, and acknowledges the
financial support of the International Max Planck Research School in
Astrophysics Ph.D. fellowship, under which this work was carried out.
GRF acknowledges the hospitality of the Max Planck Institute for
Astronomy during the inception of this work and the support of the
Departments of Physics and Astronomy, Princeton University; her
research is also supported in part by NSF-PHY-0101738.  PN
acknowledges financial support through a Zwicky Fellowship at ETH,
Zurich.

\appendix

\section*{APPENDIX A: THE COUNTS IN CELLS LUMINOSITY FUNCTION ESTIMATOR
  AND COMPARISON TO THE SWML RESULTS} 

\renewcommand{\theequation}{A\arabic{equation}}
\setcounter{equation}{0}  

\renewcommand{\thefigure}{A\arabic{figure}}
\setcounter{figure}{0}  

Our counts-in-cells (CiC) method to measure the density dependent
luminosity function and obtain its amplitude is simple and will be
illustrated with the example of a mock galaxy sample in a cubical
volume of side-length $L$.  The full luminosity function for such a
sample is trivial.  By definition it is simply the number of galaxies
in each magnitude interval divided by the volume of the box:
\begin{equation}
\Phi(M) = N(M)\ /\ L^3 ~.
\end{equation}

To determine the luminosity function as a function of local galaxy
density we require two additional pieces of information. Firstly we
sub-divide the galaxy population into density bins.  The local density
for each galaxy is calculated within an $8h^{-1}$Mpc radius as
described in Section~2.2.  This gives us the number of galaxies in
each density bin belonging to each magnitude range, $N_{\delta_8}
(M)$.

Secondly we determine the volume which should be attributed to the
various density bins.  We do this by finding the fraction of the
volume in which the galaxies of each density bin reside,
$f_{\delta_8}$.  This fraction is measured by massively oversampling
the box with \emph{randomly} placed $8h^{-1}$Mpc spheres, in each of
which we estimate a local density in the same way as before.  Once all
spheres have been placed we count the number which have a local
density in each density range.  The volume fraction of each bin is
then just the fraction of spheres found in each bin.  Since the total
volume of the box is known, the volume of each density bin is now also
known.  The density dependent luminosity function is then calculated
as:
\begin{equation}
\phi_{\delta_8} (M) = N_{\delta_8} (M)\ /\ f_{\delta_8} L^3 ~.
\end{equation}

\begin{figure}
\plotone{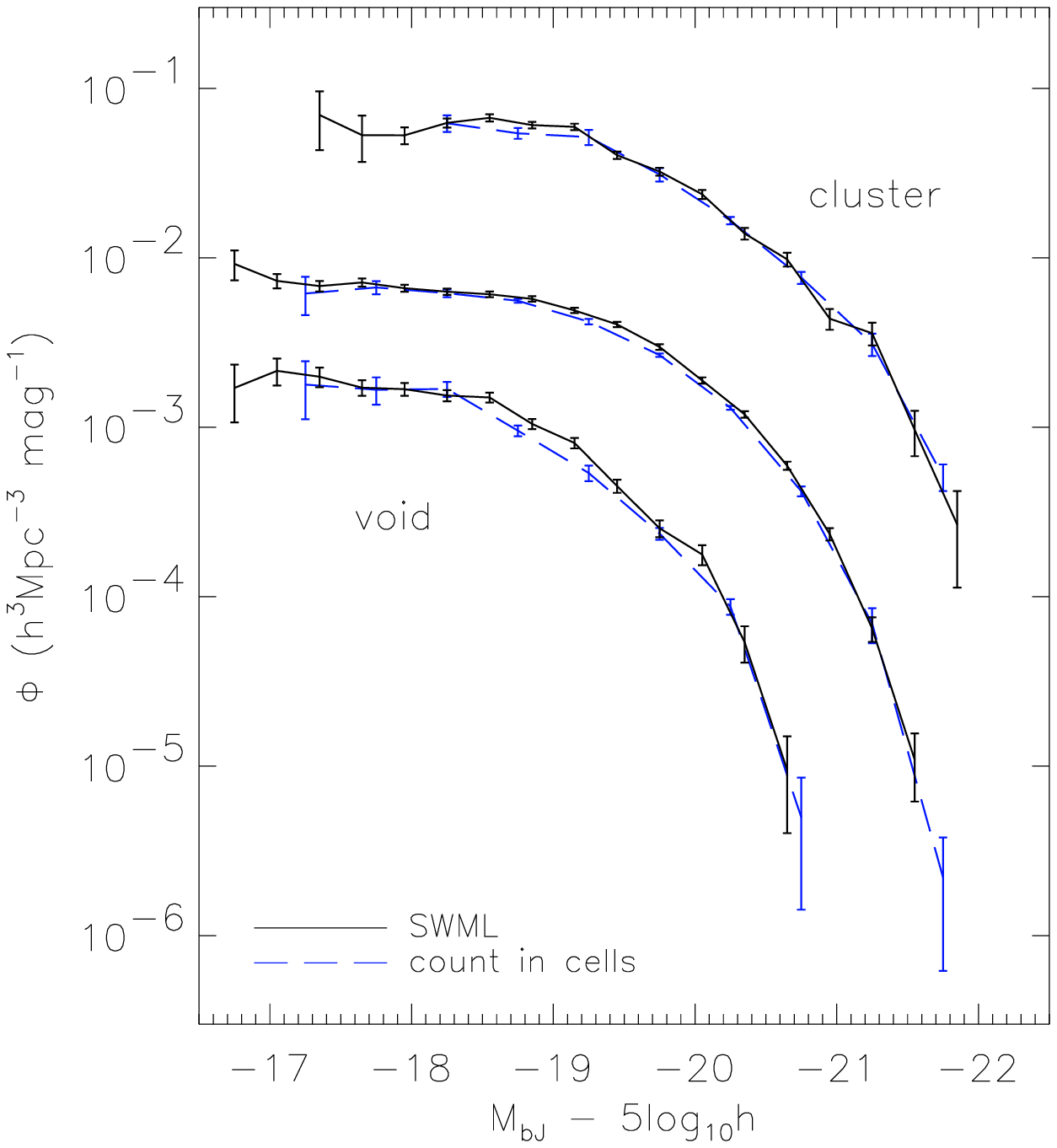}
\caption{A comparison of the raw counts in cells luminosity function
  with the normalised SWML luminosity function, as described in the
  text (Section~2.3 and Appendix~A).  Shown are the cluster, mean, and
  void populations consisting of all galaxy types only, although all
  luminosity functions used in this paper behave equally as well.
  The shapes estimated by the two very different methods are in
  very good agreement over the magnitude ranges considered.}
\end{figure}

The situation becomes more complicated when dealing with a magnitude
limited redshift survey instead of a simple simulated box.  Galaxy
counting and volume estimation must now be restricted to regions of
the survey in which the magnitude range being considered is volume
limited.  This range of course changes for each magnitude bin in which
the luminosity function is measured.  In addition, small corrections
($<10\%$) are required when counting galaxies to account for the
spectroscopic incompleteness of the survey (see Croton et al. 2004).
In all other respects, however, the calculation of $\Phi_{\delta_8}
(M)$ is the same as in the ``box'' example given above.

In Fig.~A1 we show a comparison of the 2dFGRS CiC and SWML luminosity
functions calculated from the same void, mean, and cluster galaxy
samples.  The SWML luminosity function has been normalised to the CiC
measurement as described in Section~2.2.  We see that both methods
produce almost identical luminosity distribution shapes.  This gives
us confidence that the CiC luminosity function can be used to
normalise the SWML luminosity function in an unbiased way.

Because of the volume-limited restriction of the CiC method, the
number of galaxies used to calculate the luminosity function is
smaller than for the SWML method, which draws from the larger
magnitude-limited catalogue.  However the benefit of the CiC method is
that it gives a direct measurement of the number density of galaxies
rather than just the shape of the distribution as the SWML estimator
does.  In addition, the CiC method is very easy to apply to mock
catalogues, as described above.  By combining the CiC and SWML methods
we capture the best features of both.

\section*{APPENDIX B: THE EFFECTS OF CHANGING THE DENSITY DEFINING
  POPULATION AND SMOOTHING SCALE}

\renewcommand{\theequation}{B\arabic{equation}}
\setcounter{equation}{0}  

\renewcommand{\thefigure}{B\arabic{figure}}
\setcounter{figure}{0}  

In our analysis we are required to make two important choices
before beginning.  The first is to find the widest possible absolute
magnitude range for the density defining population (DDP, see
Section~2.2) while maximising the amount of the 2dFGRS survey volume
sampled.
The second is the scale over which we smooth the DDP galaxy
distribution to determine the density contours within this volume.
We will now consider the effect of changing each of these choices in
turn.

\subsection*{The density defining population}

\begin{figure}
\plotone{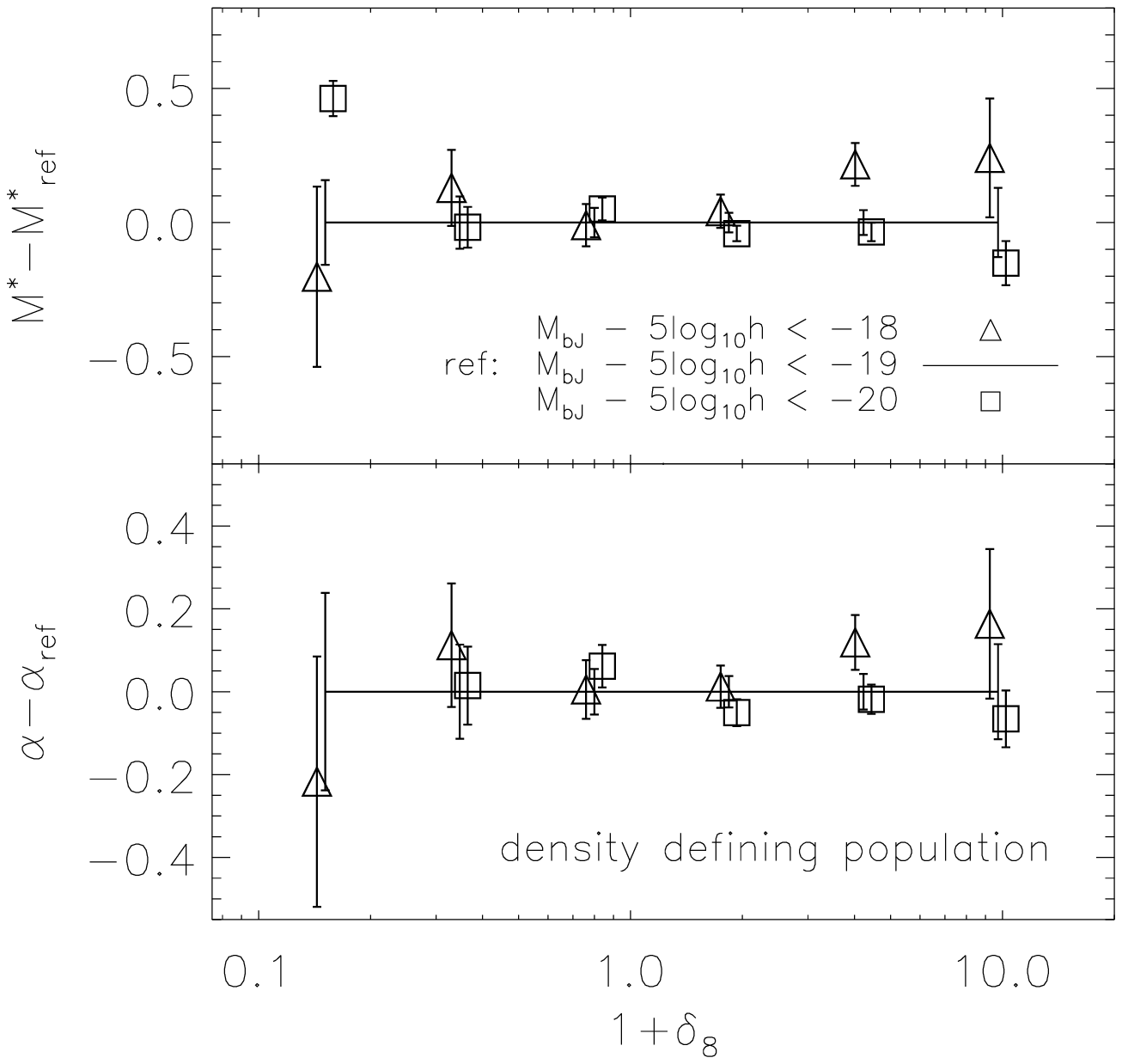}
\caption{The difference in the STY Schechter function parameters
  when the local density is calculated with an increasingly fainter
  density defining population (DDP): $M_{min}-5\log_{10}h=-20$, $-19$,
  and $-18$.  Such a change also changes the redshift range of the
  included volume as described in the text.  For clarity
  only results for all galaxy types are shown.  The reference sample
  is the $M_{min}-5\log_{10}h=-19$ DDP used throughout this paper,
  and the other DDP results are shown relative to this.}
\end{figure}

\begin{figure}
\plotone{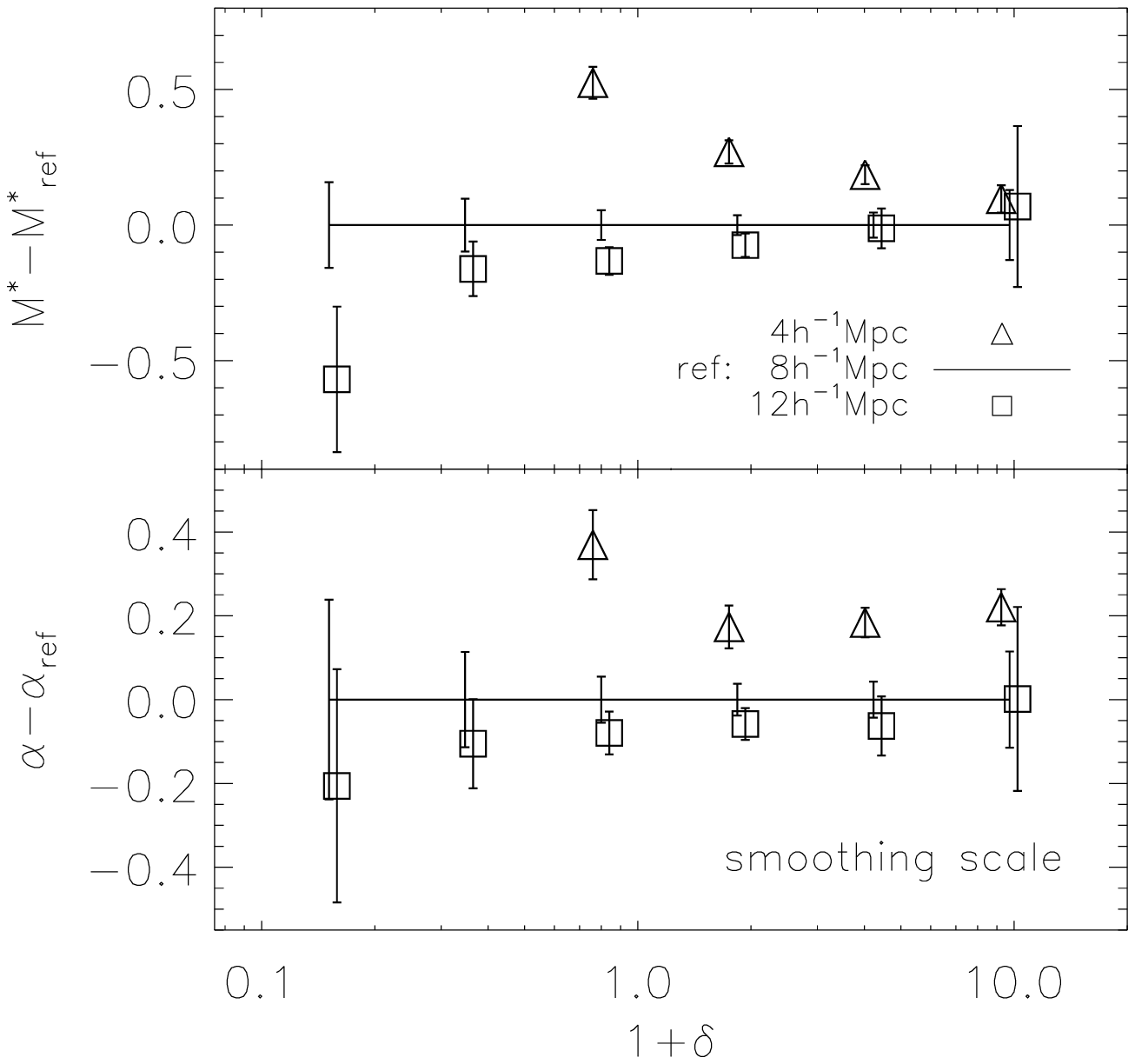}
\caption{The difference in the STY Schechter function parameters
  for different density bins when calculated by smoothing the local
  galaxy distribution on different scales: $4,8,$ and
  $12h^{-1}$Mpc. For clarity only results for all galaxy types are
  shown.  The reference sample is the $8h^{-1}$Mpc sphere smoothing
  scale used throughout this paper, and the other smoothing scale
  results are shown relative to this.}  
\end{figure}

The DDP is important in that it not only sets the mean density of
galaxies used to define the density contours, but also determines the
redshift range of the full magnitude-limited catalogue to be included
in the analysis.  Clearly one would like as high-statistics a sample
as possible in as large a volume as possible for the best results.  In
a volume limited galaxy sample such as the DDP, the maximum galaxy
redshift available is constrained by the specified faint absolute
magnitude limit: galaxies beyond this redshift range are no longer
guaranteed to be volume limited and are thus not included.  For the
DDP faint magnitude limit of $M_{min}-5\log_{10} h =-19$ the maximum
survey boundary is $z=0.13$.  Changing the faint magnitude limit to
$M_{min}-5\log_{10}h =-18~(20)$, i.e. a denser (sparser) DDP, results
in a maximum redshift boundary of $z=0.088~(0.188)$, i.e. a smaller
(larger) sampling volume.

In Fig.~B1 we show the result found when repeating the analysis of
Section~3 (Fig.~7) but using DDPs defined by different faint absolute
magnitude limits.  We plot the STY $M^*$ and $\alpha$ values for each
density bin \emph{relative} to the $M_{min}-5\log_{10} h =-19$ DDP
used throughout this paper.  The faintest DDP shown,
$M_{min}-5\log_{10} h=-18$, is approximately $8$ times denser than the
brightest, $M_{min}-5\log_{10} h=-20$, but with a volume roughly $30$
times smaller.  Even so, almost all measurements shown across all
density bins are consistent at the $1\sigma$ level, demonstrating that
our definition of the DDP is a robust representation of the underlying
global density distribution.

\subsection*{The smoothing scale}

Now let us examine how changing the smoothing scale with which we
define local density affects the shape of our luminosity functions. In
Fig.~B2 we examine the values of the Schechter parameters when
measured with spheres of radius $4$ and $12h^{-1}$Mpc, compared to
when the luminosity function is measured with an $8h^{-1}$Mpc sphere.

Fig.~B2 shows a typical deviation of $<0.2$ magnitudes for $M^*$ and
$<0.2$ for $\alpha$.  The $4h^{-1}$Mpc smoothing scale deviates
strongly from the other values in the under-dense regions (the first
two points lie beyond the axis range plotted), however in these
environments such a smoothing scale gives a poor estimate of the local
galaxy density due to Poisson noise in small number counts.  
Indeed, Hoyle et al. (2004) have shown that in the extreme under-dense
2dFGRS survey regions the characteristic scale of voids is
approximately $15 h^{-1}$Mpc. 
For cluster regions $4h^{-1}$Mpc spheres can be employed and would give a
higher resolution discrimination of the structure.  
The $8h^{-1}$Mpc smoothing scale we have adopted captures the essential
aspects of voids while roughly optimising the statistical signal, and
is thus a good probe of both the under and over-dense regions of the
survey volume.

\section*{APPENDIX C: SYSTEMATIC EFFECTS WHEN ESTIMATING THE SCHECHTER
  FUNCTION PARAMETERS}

\renewcommand{\theequation}{C\arabic{equation}}
\setcounter{equation}{0}  

\renewcommand{\thefigure}{C\arabic{figure}}
\setcounter{figure}{0}  

One may ask to what degree the trends seen in Fig.~7 and
Table~1 are influenced by the systematics discussed in Section~2.4.
We note there that our STY measurements recover a flatter faint end
than the current published 2dFGRS luminosity function for the
completed catalogue.  We identify three systematic effects which
contribute to this behaviour: the absolute magnitude range considered
when applying the STY estimator, the fact that the luminosity function
is not perfectly described by a Schechter function, and the
sensitivity of the faint-end slope parameterisation to model-dependent
corrections included to account for missed galaxies.

We find that the first two of these effects have the strongest
influence on the measured STY faint-end slope. Indeed, testing the
first reveals that any STY estimate of the 2dFGRS luminosity function
over a restricted absolute magnitude range displays a systematic shift
in the recovered STY parameters along a line in the $M^*-\alpha$
plane. The brighter the faint magnitude restriction, the flatter the
faint-end slope is and the fainter the characteristic magnitude
becomes.  Such behaviour is a consequence of small but important
deviations in the galaxy luminosity function shape from the pure
Schechter function assumed by the STY estimator. In addition, Fig.~11
of Norberg et al. (2002) reveals a dip in the luminosity function
between the magnitude range $-17 > M_{\rm{b_J}} - 5 \log_{10} h > -18$
and a steepening faintward of this. In our analysis only galaxies
brighter than $M_{\rm{b_J}} - 5 \log_{10} h = -17$ are considered due
to the restriction of the DDP.  This limitation adds extra weight to
the influence of the dip on the STY fit contributing further to a
flatter estimation of $\alpha$. When mock galaxy catalogues
constructed to have a perfect Schechter function luminosity
distribution are analysed in an identical way to the 2dFGRS samples,
we find that the above systematics all but disappear and the ``true'' $M^*$
and $\alpha$ values are recovered for any reasonable choice of STY
fitting range.

Finally, we note that the sensitivity of the faint-end slope
parametrisation to systematic corrections for spectroscopically missed
galaxies is minimised by restricting our analysis to galaxies with
$\rm{b_J} < 19$, for which the spectroscopic incompleteness is
typically less than $\sim 8$\% (see Fig.~16 of Colless et al. 2001).

Given that a full correction of the above systematic effects is not
possible in our analysis, the next best thing we can do is try to
quantify to what degree they influence our results and conclusions.
We do this by fixing the faint-end slope $\alpha$ when applying the
STY estimator: at $-1.2$ for the all-types samples, $-1.1$ for the
late-type samples, and at $-0.5$ for the early-type samples.  Such
choices enforce the published field luminosity function faint end
values found by Norberg et al. (2002a) and Madgwick et al. (2002) and
remove the degeneracy in the $M^*-\alpha$ plane.

\begin{figure}
\plotone{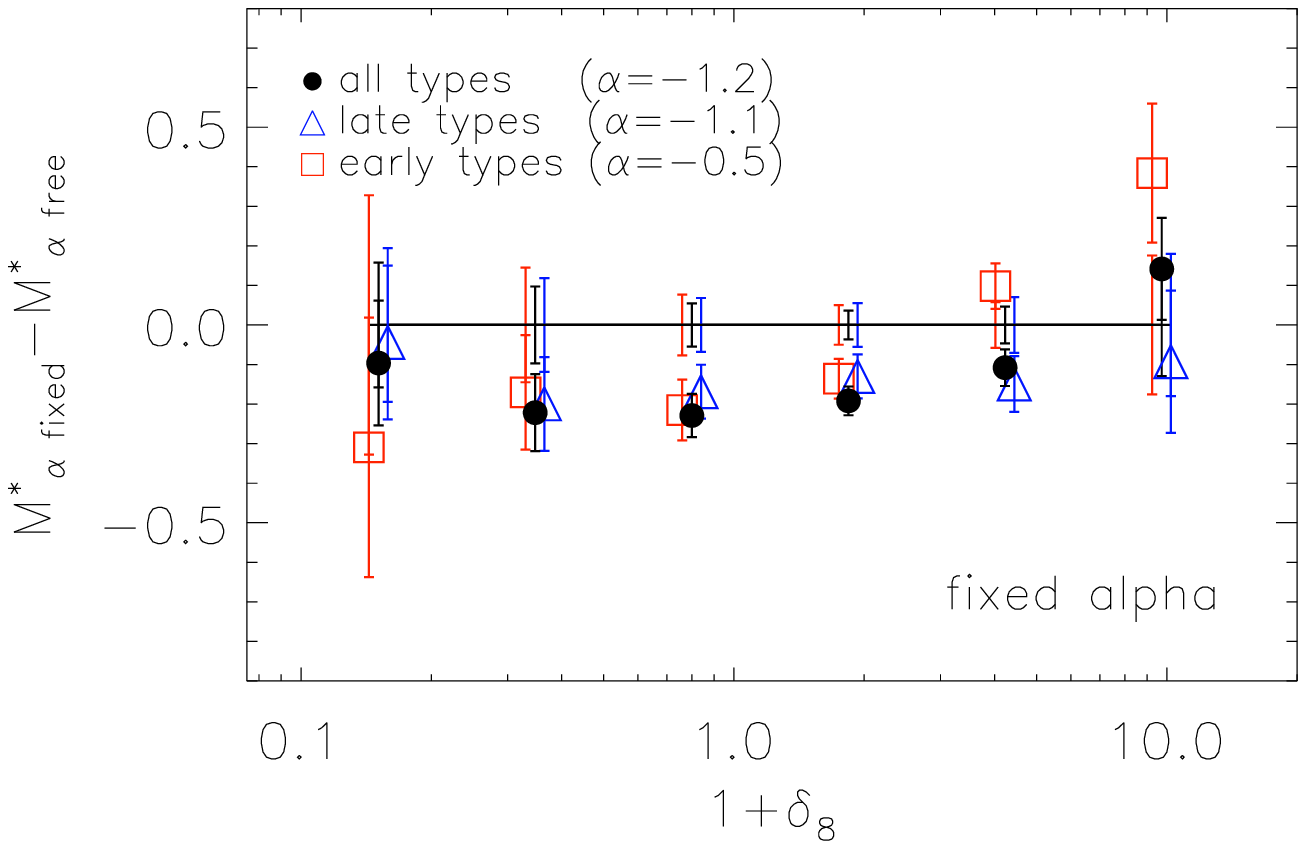}
\caption{The shift in the STY Schechter function parameter $M^*$ when
  $\alpha$ is kept fixed at the published field value, compared with
  that found when $\alpha$ remains free (Table~1 and Fig.~7).  The
  points are plotted as a function of local density and shown for each
  galaxy type and the combined all-type sample. 
}
\end{figure}

Fig.~C1 shows the size of the shift in $M^*$ when such constraints are
applied relative to that found when $\alpha$ is allowed to remain free
(i.e. Table~1 and Fig.~7).  Most notable here is that, apart from the
two most over-dense bins in the early-type sample, there is no
significant difference in the {\em behaviour} of $M^*$ with density
environment.  The approximate $0.2$ magnitude offset seen in this
figure can be understood by remembering that because the faint-end
slope we measure when $\alpha$ remains free is slightly flatter than
the published values (due to the systematics discussed above), by
artifically fixing $\alpha$ one forces $M^*$ to move to compensate.
The important point is that the {\em trends} seen in Fig.~7 with
changing local density remain unchanged.

For the two most over-dense early-type samples a shift of up to $0.4$
magnitudes is seen.  We note from Table~1 that our best-fit (free
$\alpha$) early-type cluster value of $\alpha = -1.12 \pm 0.14$ is
well matched by the equivalent 2dFGRS De Propris et al. (2003) result
of $-1.05 \pm 0.04$.  In effect, by constraining the early-type
cluster faint-end slope to the field value of $-0.5$ we ignore the
real changes in galaxy population seen between the Madgwick et
al. (their Fig.~10) and De Propris et al. (their Fig.~3) luminosity
functions (see also our Fig.~2).  Such population changes, we argue,
result in the strikingly different Schechter function parameterisation
behaviour seen in Fig.~7 and Table~1 for early and late-type galaxies.
Fig.~C1 gives us confidence that the $M^* - \alpha$ degeneracies and
systematics investigated here are not significantly influencing our
results or the conclusions we draw from them.

\label{lastpage}

\end{document}